\documentstyle[a4]{article}

\input mssymb.sty

\begin{document}
\topmargin=-2cm
\def\ss{\subset}
\def\si{\sigma}
\def\bw{\bigwedge}
\def\ra{\rightarrow}
\def\al{\alpha}
\def\ba{\beta}
\def\om{\omega}
\def\lb{\lambda}
\def\ga{\gamma}
\def\Si{\Sigma}
\def\Lb{\Lambda}
\def\be{\begin{equation}}
\def\ee{\end{equation}}
\def\cn{{\cal C}^\infty}
\def\rc{{\Bbb R}\,}
\def\ti{\tilde}
\def\pd{\partial}
\def\Ch{{\frak{Ch}}}
\def\Ph{{\frak{Ph}}}
\input epsf

\begin{center}
\vbox{}
\vskip 1cm
{\LARGE Contact geometry in Lagrangean mechanics\\}
\vskip 1cm
{\large Pavol \v Severa\\}
\vskip 0.5cm
{\small Dept. of Theor. Physics,
Charles University,\\ \small V Hole\v sovi\v ck\'ach 2, 18000 Prague, Czech
Republic}
\vskip 1cm
\end{center}

\begin{abstract}
We present a picture of Lagrangean mechanics, free of some unnatural features
(such as complete divergences). As a byproduct, a completely natural
$U(1)$-bundle over the phase space appears. The correspondence between
classical and quantum mechanics is very clear, e.g. no topological
ambiguities remain. Contact geometry is the basic tool.
\end{abstract}

\section{Introduction}

In this paper we show how to get rid of some unnatural features of
Lagrangean mechanics, such as multivaluedness and neglecting
total divergences. There is almost nothing new: we simply consider
Hamilton--Jacobi equation and its characteristics. The only point is
in introducing, instead of $M\times\rc$, a principal $G$- bundle $U$
over the spacetime $M$, where $G=\rc$ or $U(1)$.
Even if $U$ is trivial, it is, in a natural way, only
a bundle and not a product. This correspods to ``up to a total divergence''
phrases. The Hamilton--Jacobi equation is simply a $G$-invariant hypersurface
in the space of contact elements of $U$.

In quantization, the wave functions are sections
of a line bundle associated to $U$, no topological ambiguity remains, so the
correspondence classical $\leftrightarrow$ quantum is very clear. The
space of characteristics $\Ch$ carries a natural contact structure; the
phase space $\Ph$ emerges as the quotient of $\Ch/G$. Thus
$\Ch\ra \Ph$ is a principal $U(1)$- (or $\rc$- ) bundle; the contact structure
gives us a connection. 

The plan of the paper is as follows:
In Section 2 we present basic facts of contact geometry, its connection
with symplectic geometry and geometrical quantization,
with first order PDE and the method
of characteristics and with asymptotics of linear PDE. In Section 3
we introduce the point of view described above and discuss its correspondence
with Lagrangians. For example, it may contain some additional topological
information (obviously the topological quantization
ambiguity has to be hidden somewhere).
The bundle $\Ch\ra\Ph$ and quantization are discussed in Section 4.
We conclude with the fact that one can replace the group $U(1)$ by any
Lie group almost without changing anything. Finally we mention
 the obvious open problem --
what happens if we do not consider extremal curves, but surfaces etc.

\section{Basic notions of contact geometry}

A {\em contact structure} on a manifold $M$ is a field of hyperplanes
$HM\ss TM$
(a subbundle of codimension 1) satisfying a maximal nonintegrability condition.
It can be formulated as follows: as for any subbundle of $TM$, we have a map
$\si:\bw^2HM\ra TM/HM$ satisfying (and defined by) the fact that for any
1-form $\al$ on $M$, annulated on $HM$, the formula
$$ \al\left(\si(u,v)\right)=d\al(u,v) $$
holds for any $u,v\in H_xM$, $x\in M$. Alternatively, we may extend $u$
and $v$ to
sections of $HM$; their commutator at $x$ (when considered mod $HM$)
is $\si(u,v)$. The maximal nonintegrability condition requires $\si$ to be
regular. In that case, $M$ is clearly odd-dimensional. Any two contact
manifolds with the same dimension are locally isomorphic (a form of Darboux
theorem).

We call a vector field on $M$ {\em contact}, if its flow preserves the contact
strucure. There is a 1-1 correspodence between contact vector fields and
sections of the line bundle $TM/HM$. More precisely, for any $w\in\cn(TM/HM)$
there is a unique contact $v$ that becomes $w$ when considered mod $HM$.
The proof is easy: choose any $v'$ that is $w$ mod $HM$. As a rule, $v'$ is not
contact, so it generates an infinitesimal deformation of the contact structure
-- say $\ba:HM\ra TM/HM$. But due to the nondegeneracy of $\si$ there is 
a unique $v''\in\cn(HM)$ producing the same deformation. Thus $v=v'-v''$ is 
the required contact field. The field $w$ is called the {\em contact
hamiltonian} of $v$.

An important example of contact geometry emerges when $M$ is a principal
$G$-bundle over a symplectic manifold $(N,\om)$, where $G=\rc$ or $U(1)$.
Suppose we are given a connection on $M$ such that its curvature is $\om$.
The horizontal distribution makes $M$ into a contact manifold. We can use
the connection 1-form $\al$ to identify sections of $TM/HM$ (contact
hamiltonians) with functions on $M$. The local flow generated by a contact
field $v$ preserves the structure of $G$-bundle iff $v$ is
$G$-invariant, i.e. iff its contact hamiltonian $f$ is (the pullback of)
a function on $N$. Then the field $v$ is projected onto a well-defined
vector field $v_N$ on $N$ whose flow preserves $\om$; in fact, $f$ is a
hamiltonian generating $v_N$. We may put these facts together:
The Lie algebra $\cn(N)$ (with the Poisson bracket) is isomorphic to the Lie
algebra of $G$-invariant contact fields on $M$. A function $f$ on $N$
and the corresponding hamiltonian vector field $v_N$ are combined together
($f$ as the vertical part and $v_N$ as the horizontal part) to form a contact
field $v$ on $M$.

This point of view is useful in geometrical quantization. Here one considers
a line bundle $L\ra N$ associated to $M\ra N$, and represents the Lie algebra
$(\cn(N),\{,\})$ by operators on the space $\cn(L)$. The sections of $L$ are
simply functions on $M$ equivariant with respect to $G$ and the action of
a function $f\in\cn(N)$ on such a section is given by the derivative with
respect to the corresponding contact vector field.

The classical example of a contact manifold  is the space of contact elements
(i.e. hyperplanes in the tangent space) of a manifold $M$, which we denote as
$CM$. The distribution $H(CM)$ is given as follows: take an $x\in CM$; it 
corresponds to a hyperplane $H$ in $T_{\pi(x)}M$, where $\pi:CM\ra M$ is the
natural projection. Then $H_x(CM)$ is $(d_x\pi)^{-1}(H)$.

Contact geometry, in particular on $CM$, was invented to give a geometrical
meaning to first order partial differential equations and to Lagrange method
of characteristics. Suppose $E\ss CM$ is a hypersurface; it will represent
the equation. Any hypersurface $\Si\ss M$ can be lifted to $CM$: for any point
$x\in \Si$ take the hyperplane $T_x\Si$ to be a point of the lift $\ti\Si$.
$\ti\Si$ is a Legendre submanifold of $CM$, i.e. $T\ti\Si\ss H(CM)$ and 
$\ti\Si$ has the maximal possible dimension
(${\rm dim}\,CM=2\,{\rm dim}\,\ti\Si+1$). $\Si$
is said to solve the equation if $\ti\Si\ss E$. This has a nice interpretation
due to Monge: For any $x\in M$ we take the enveloping cone of the hyperplanes
$\pi^{-1}(x)\cap E$ in $T_xM$. In this way we obtain a field of cones in $M$.
Then $\Si$ solves the equation if it is tangent to the cones everywhere.

Lie's point of view is to forget about $M$ and to take as a solution any
Legendre submanifold contained in $E$. Such a solution may look singular in $M$
(singularities emerge upon the projection $\pi:CM\ra M$). This definition uses
only the contact structure on $CM$ and thus allowes using the entire
(pseudo)group
of contact transformations.

Now we will describe the method of characteristics. The hyperplane field
$H(CM)$ cuts a hyperplane field $HE$ on $E$ (there may be points where the
contact hyperplane touches $E$. Generally they are isolated and we will ignore
them). The field $HE$ does not make $E$ into a contact manifold: the form $\si$
becomes degenerate when we restrict ourselves from $H_x(CM)$ to $H_xE$.
Thus at any $x\in E$ there appears a direction along which $\si$ is
degenerate. The integral curves of this direction field are called 
{\em characteristics}. For example, if the Monge cones coming from  $E$
are the null cones of some pseudo-riemannian metrics on $M$ then
the projections
of the characteristics are the light-like geodesics in $M$. 

Generally, if $F$ is a manifold with a hyperplane field $HF$, and the form
$\si:\bw^2HF\ra TF/HF$ has constant rank, then the bundle of kernels of $\si$,
$KF\ss HF$, is integrable. Moreover, if one takes  an open $U\ss F$ small
enough, so that the integral manifolds of $KF$ in $U$ form a manifold $\Ch$,
then there is a well-defined contact structure on $\Ch$ coming from the 
projection of $HF$. Coming back to the case of $E\ss CM$, it gives us a method
of finding the Legendre submanifolds contained in $E$. Just take a submanifold
that is almost Legendre -- up to the dimension, which is less by 1. Suppose
that the characteristics intersect it transversally. Then their union form
a Legendre submanifold.

Let us look at vector fields on $E$ with flow preserving the field $HE$;
we shall call them contact, too.
First of all, there are {\em characteristic vector fields}, i.e. fields
touching the characteristics. Thus it is no longer true that if we choose
a $w\in\cn (TE/HE)$ then there is a unique $v\in\cn (TE)$ equal to $w$ 
mod $HE$: we can always add a characteristic field to $v$. On the other hand,
$w$ cannot be arbitrary. The flow of a contact field has to preserve the
characteristic foliation. If $\Ch$ is the space of characteristics, each
contact field on $E$ can be projected onto a contact field on
$\Ch$ (recall $\Ch$ is a contact manifold). This is the basis for conservation
laws.
For example if a contact field $v\in HE$ (i.e. $w=0$)
at a point $x\in E$ then
$v\in HE$ ($w=0$)
along  the characteristic $\ga_x$ running through $x$.
Let us also notice that any contact vector field on $E$ can
be prolongated to a contact vector field on $CM$
(with the flow preserving $E$).

Hypersurfaces $E\ss CM$ often come from an equation of the type $Df=0$,
where $D:\cn (M)\ra\cn (M)$ is a linear differential 
operator. Take the sybmol $s_D$ of $D$ (a function on $T^*M$ defined by
$(i\lb)^n s_D(dg)=D\exp(i\lb g)+O(\lb^{n-1})$, $\lb\ra\infty$, where $n$ is the
degree of $D$ and $g\in\cn(M)$).
The equation $s_D=0$ specifies a hypersurface $E\ss CM$.
The singularities of solutions of $Df=0$ are located on hypersurfaces solving
the equation corresponding to $E$; also, if $f=a(x)\exp(i\lb S(x))$, 
$\lb\ra\infty$ is an asymptotic solution of $Df=0$ then the levels 
$S(x)=const$ solve the $E$-equation. 

\section{The geometry of Lagrangean mechanics}

We shall deal with first-order variational principles. Suppose that at each
point $x$ of a manifold $M$ (the space-time or extended configuration space)
there is a 1-homogeneous function $\Lb_x:T_xM\ra\rc$ (and suppose everything
is smooth outside the zero section of $TM$). Then on each oriented curve $\ga$,
$\Lb$ specifies a 1-form, so we may compute its integral $S(\ga)=\int_\ga\Lb$.
We are looking for extremals of $S$ (in this paper, extremal means
stationary).

There are several reasons why this point of view is not entirely satisfactory.
First of all, even in the simplest problems, $\Lb_x$ is not defined on all
$T_xM$, but only on an open conic subset. Even worse, $\Lb$ may be multivalued.
An example is drawn on the following two figures. On the first one, we suppose
that $\Lb_x$ is positive (outside 0). The figure represents the endpoints of
vectors satisfying $\Lb_x(v)=1$; it is called the {\em wave diagram} in the
beautiful elementary book \cite{burke}. The dashed lines represent a covector
$p$ corresponding to the drawn vector (they are $p=0$ and $p=1$); $p$ is
called the {\em momentum}. 

$$\epsfxsize 35mm
\epsfbox{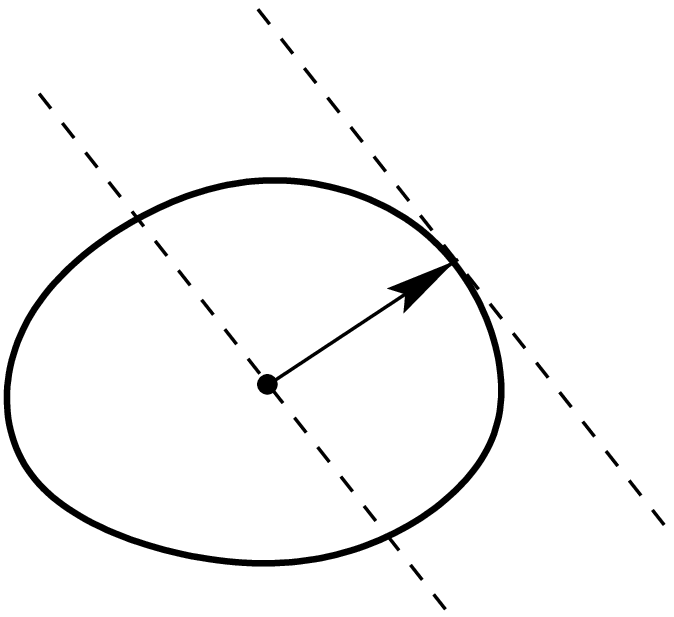 }$$
Obviously, we may use the field of wave diagrams instead of $\Lb$. But we
may work as well with diagrams of the following shape; they correspond to
multivalued $\Lb$'s:

$$\epsfxsize 30mm
\epsfbox{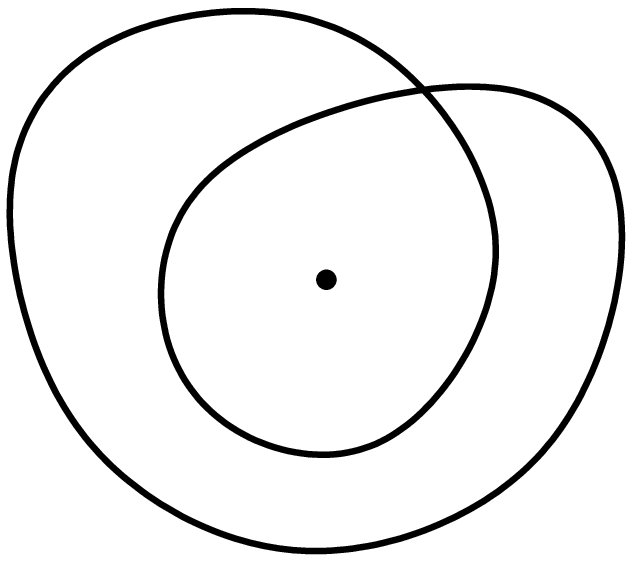}$$

However, the real problem is that $\Lb$ is unnatural. The reason is that it is 
defined only up to a closed 1-form. For example, in the presence of an
`electromagnetic
field' $F\in\cn (\bw^2 T^*M)$, $dF=0$, we take
as the actual $\Lb$ (the one from which we compute $S$) $\Lb+A$,
where $dA=F$. Of course $A$ need not exist globally and it is not
defined uniquely.

This problem appears also in Noether theorem: we take as an infinitesimal 
symmetry any vector field $v$ whose flow preserves $\Lb$ up to some $df$.
It is desirable to have a picture in which $v$ is an actual symmetry.

A way out is in the following construction: Let $U\ra M$ be a principal
$G$-bundle, where $G=U(1)$ or $\rc$ (you may imagine that we added the action
$S$ to $M$ as a new coordinate; of course this interpretation is rather 
limited). Suppose we are given a $G$-invariant hypersurface $E\ss CU$;
we are interested in its characteristics. Their projections to $M$ are the
extremals for certain (multivalued) $\Lb$ (if $c_1(U)\ne0$
then either $\Lb$ exists only locally or we must admit an elmg. field $F$).
We simply replaced $\Lb$ by the corresponding Hamilton--Jacobi equation $E$,
but the new point of view is rid of the problems listed
above. {\em For this reason we take $E\ss CU$ and its characteristics
as fundamental
and the Lagrangian $\Lb$ as a derived, sometimes ill-defined notion.}

The correspondence between $E$ and $\Lb$ is as follows: Let $\al$ be an
arbitrary connection
1-form on $U$. To find the wave diagram at a point $x\in M$, take a point
$y\in U$ above $x$. The intersection of the Monge cone in $T_yU$ with the
hyperplane $\al=1$ is the wave diagram. We have to take the curvature $F$
as the elmg. field. We see that the transformation $\Lb\ra\Lb+A$, $F\ra F-dA$
($A$ a 1-form) corresponds
simply to a change of the connection.

If we start with $\Lb$ and $F$, we have to suppose that
the periods of $F$ are integral (or at least commesurable) to find a $U$
admitting a connection with $F$ as the  curvature. Notice that
if $H^1(M,G)\ne 0$, the picture
$E\ss CU$ contains
more information than the pair $(\Lb,F)$ .
The inequivalent choices of $U$ together with a connection correspond
to the elements of the group $H^1(M,G)$ (this group acts there freely and
transitively). The subgroup $H^1(M,{\Bbb Z})\otimes G$ corresponds
to equivalent $U$'s (with ineqivalent connections); if $G=U(1)$, even
the quotient
group may be nontrivial (it is ${\rm Tor}\,H^2(M,{\Bbb Z}$)). These ambiguities
are clearly connected with quantization.

A well known example is the following: Let the Monge cones on $U$ be the
light cones of a Lorentzian metrics and suppose the vector field $u_G$ 
generating the action of $G$ is spacelike. As a connection on $U$ take the
orthogonal complements of $u_G$. Then the wave diagrams are the (pseudo)spheres
of a Lorentzian metrics on $M$. This picture describes a charged relativistic
particle and its antiparticle in an elmg. field given by the curvature of the
connection.\footnote{The connection dissects each light cone in $U$ into two
halfs. Thus the lightlike geodesics in $U$ (the characteristics) are (at least
locally, and globally if there is a time orientation) divided into 3 classes;
two of them are projected onto particles and
antiparticles worldlines respectively, while the curves in the third class are
horizontal and they are projected onto lightlike geodesics in $M$.} In the
nonrelativistic limit the field $u_G$ becomes lightlike and the antiparticle
disappears.

Let us look at Noether theorem. In the $(\Lb,F)$-picture
one takes as a symmetry a vector field $v$ together with a function $f$
satisfying
$$v(\Lb)+F(v,.)+df=0$$
($v(.)$ denotes the Lie derivative); then $p(v)+f$ is constant on extremals.
But for $E\ss CU$ we simply take a $G$-invariant vector field on $U$ 
preserving $E$. In fact one easily sees the full statement of Noether theorem,
claiming a 1-1 correspondence between conservation laws and $G$-invariant
contact fields
on $E$ modulo characteristic fields.


\section{A U(1)-bundle over the phase space and quantization}  

Let us suppose that the characteristics in $E$ form a manifold $\Ch$.
It inherits a contact structure. Notice
that $E$ is a $G$-bundle; we shall also suppose that the group $G$ acts nicely
on $\Ch$ so that $\Ch$ becomes a $G'$-bundle where $G'=G/H$ and $H\ss G$ is 
discrete. Its base $\Ph=\Ch/G'$ is the phase space.
Where the contact hyperplanes
on $\Ch$ may be used as a connection for $\Ch\ra \Ph$, the curvature is
the usual symplectic form on $\Ph$. The points of $\Ph$ where this is
impossible
are usually deleted and they should be regarded as ideal. For example,
the full $\Ph$ of a relativistic particle in 1+1-dimensions is on the following
picture:   

$$\epsfxsize 25mm
\epsfbox{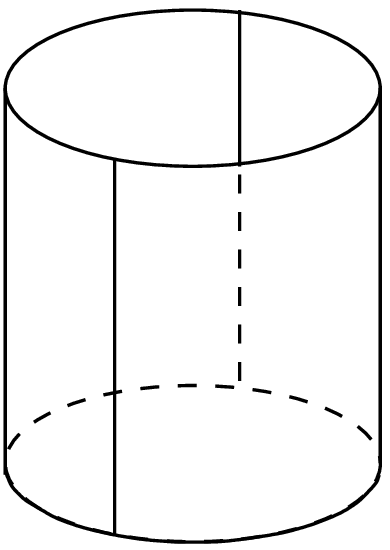}$$
One half of the cylinder corresponds to particles, the other half to
antiparticles and the connecting lines to lightlike geodesics. 

We see that there is a completely natural $U(1)$- or $\rc$-bundle $\Ch$ over
the phase space, together with a natural connection. It is important in the
view of use of such a bundle in quantization. Notice that $\Ch$ is even prior
to $\Ph$.

Let us now look at quantization using wave functions in $M$. This may have
nothing to do with quantum mechanics: we simply look for a wave equation
that leads to a given classical picture in a limit. Usually, one considers
linear equations $D_hf=0$ ($h$ being a parameter in $D$) and looks for the
high-frequency asymptotics as $h\ra0$ and the wavelength is of order
$h$. It is however much nicer if $D$ is fixed; an outline of the theory
was given at the end of Section 2. Thus let $D$ be a $G$-invariant
linear diff. operator on $U$. If we consider only $G$-equivariant functions 
(with the weight $1/h$), we get an operator $D_h$ on the corresponding
associated bundle. 

For example, the Schroedinger equation comes from
$$\left({1\over 2m}\triangle+V(x,t){\pd^2\over\pd s^2}+{\pd^2\over\pd s\pd t}
\right)\psi(x,t,s)=0,$$
where $s$ is the new coordinate (here $U=M\times\rc$):
just notice that $\pd/\pd s$ becomes $i/\hbar$ for $\psi$ with the weight
$1/\hbar$.

Let $E\ss CU$ be given by $s_D=0$ where $s_D$ is the symbol of $D$
(notice that the Monge cone in $T_xU$
is dual to the cone $s_{D,x}=0$ in $T^*_xU$).
In the obvious sense the equation $D_hf_h=0$ gives the classical $E$-theory
as $h\ra 0$. For example, take a (nonequivariant!) solution of $Df=0$ with
a singularity on a narrow strip along a characteristic of $E$. If we take
the Fourier component $f_h$ for $h\ra0$, it is significantly non-zero only
close to the projection of the characteristic to $M$. Perhaps an interesting
point is that the equation $Df=0$ contains $D_hf_h=0$ for any $h$.

Thus given $E$, quantization simply means a $G$-invariant $D$ giving $E$
by $s_D=0$. Of course, the Monge cones of $E$ have to be algebraic.

Finally, let us return to $\Ch\ra \Ph$. We have a situation typical
to integral geometry: $\Ch\leftarrow E\ra U$. In geometrical quantization
one considers sections of bundles associated to $\Ch\ra \Ph$, but here we take
all possible $h$'s at once, so we consider all the functions on $\Ch$ instead.
One should expect a correspondence between certain such fuctions and 
functions on $U$ satisfying $Df=0$. A polarization on $\Ph$ gives us
a $G$-invariant Legendrean foliation (if it is real) or (if it is completely
complex) a  $G$-invariant
(codimension 1
and nondegenerate) $CR$-structure  on $\Ch$.
The foliation gives us a complete system of solution of the Jacobi--Hamilton
equation. Thus functions on $\Ch$, constant on the leaves of the foliations,
should correspond to solutions of $Df=0$ that are (integral) linear
combinations of functions singular along hypersurfaces in the complete system.
The $CR$-case is somewhat more complicated.

The discussion above is useless in this complete generality
(and several important points were omitted), but it might be interesting
for some classes of $D$'s.

\section{Conclusion}
In the present paper $G$ was always 1-dimensional, but one can consider a
principal $G$ bundle $U\ra M$ and a hypersurface $E\ss CU$ for another
Lie group $G$. The manifold $\Ch$ is still contact, but $\Ph=\Ch/G$ is no
longer symplectic; it carries only an analogue of symplectic structure.
Characterictics of $E$ represent particles in a Yang--Mills field.
We can also consider a $G$-invariant operator $D:\cn(U)\ra\cn(U)$. Suppose
$V$ is a $G$-module and the dual $V^*$ contains a cyclic vector $\al$.
Let $I$ be the ideal in $U({\frak g})$ of elements annulating
$\al$. Then we can
embed $V$ into the regular representation (namely onto functions annulated by
$I$) via $v\mapsto\al(gv)$. In this way the functions on $U$ annulated by
$I$ are sections of the vector bundle associated to $V$. Thus $D$
becomes an operator on these sections. We see the situation is quite analogous
to 1-dimensional $G$.

Perhaps the real problem is to go from extremal curves to surfaces and higher.
The problems with Lagrangians remain the same.


\subsection*{Acknowledgement}
This work was partially supported by the grant GA\v CR 201/96/0310.

\end{document}